\documentclass[pre,twocolumn,superscriptaddress,showpacs]{revtex4-1}

\usepackage{graphicx}
\usepackage{amsmath}
\usepackage{latexsym}
\usepackage{comment}
\usepackage{bm}
\usepackage{gensymb}
\usepackage{hyperref}
\usepackage{color}
\usepackage{colortbl}
\usepackage[font=footnotesize,labelfont=bf,
   justification=centerlast ]{caption} 
\usepackage[range-phrase=-]{siunitx}
\hypersetup{colorlinks,urlcolor=blue}

\renewcommand{\vec}[1]{\bm{#1}}

\definecolor{Gray1}{gray}{0.88}
\definecolor{Gray2}{gray}{0.98}

\graphicspath{{Figures/}}

\begin{document}
\title{Simulation and observation of line-slip structures in columnar structures of soft~spheres}
\date{\today}
\author{J. Winkelmann} 
\affiliation{School of Physics, Trinity College Dublin, The University of Dublin, Ireland}
\author{B. Haffner} 
\affiliation{School of Physics, Trinity College Dublin, The University of Dublin, Ireland}
\author{D. Weaire} 
\affiliation{School of Physics, Trinity College Dublin, The University of Dublin, Ireland}
\author{A. Mughal}
\affiliation{Institute of Mathematics, Physics and Computer Science, Aberystwyth University, Penglais, Aberystwyth, Ceredigion, Wales, SY23}
\author{S. Hutzler} 
\affiliation{School of Physics, Trinity College Dublin, The University of Dublin, Ireland}

\begin{abstract}
We present the computed phase diagram of columnar structures of soft spheres under pressure, of which the main feature is the appearance and disappearance of line slips, the shearing of adjacent spirals, as pressure is increased. A comparable experimental observation is made on a column of bubbles under forced drainage, clearly exhibiting the expected line slip.
\end{abstract}

\maketitle


\section{Introduction}

Columnar structures are ubiquitous throughout biology: examples range from viruses, flagella, microtubules, microfillaments, as well as certain rod shaped bacteria \cite{erickson1973tubular, hull1976structure, brinkley1997microtubules, bryan1974microtubules, amir2012dislocation}. However, they are also increasingly recognised in the physical sciences, particularly in nano-physics \cite{smalley2003carbon, chopra1995boron, brisson1984tubular} in the form of nanotubes and nanowires. Other notable examples include the helical self-assembly of dusty plasmas \cite{tsytovich2007plasma}, asymmetric colloidal dumbbells \cite{zerrouki2008chiral} and particles trapped in channels \cite{yin2003self}. More recently such columnar arrangements have also been found as components of exotic crystal structures \cite{douglass2017stabilization, wang2016electric}.

\begin{figure}
\begin{center}
\includegraphics[width=0.9\columnwidth]{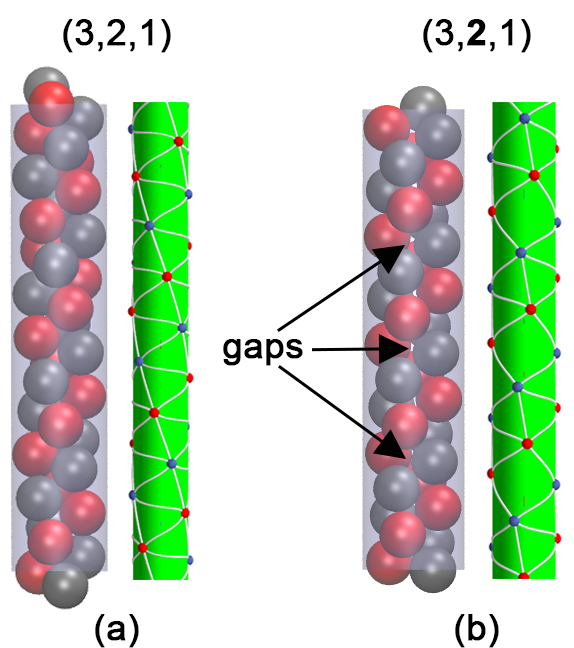}
\caption{Examples of columnar hard sphere structures. For each pair of images the first shows the arrangement of hard spheres inside a tube, while the second shows a skeleton diagram in which sphere centres are represented by points and contacts between spheres are indicated by a line joining the sphere centres. The structures shown are (a) a uniform arrangement $(3, 2, 1)$; and, (b) a related line-slip structure $(3, \bm{2}, 1)$.
The gaps in (b) correspond to a loss of contacts compared to (a) - the loss of contacts can also be seen clearly in the skeleton diagrams.}
\label{bubblespheres}
\end{center}
\end{figure}

Columnar structures arise in their most elementary form when we seek the densest packing of hard spheres inside (or on the surface) of a circular cylinder \cite{mughal2011phyllotactic, chan2011densest,  mughal2012dense, mughal2013screw, mughal2014theory, yamchi2015helical,fu2016hard}. A wide range of structures have been identified and tabulated \cite{mughal2012dense}, depending on the ratio of cylinder diameter $D$ to sphere diameter $d$. For each of a set of discrete values of $D/d$, a uniform structure is found, and may be labelled with the traditional phyllotactic indices $(l,m,n)$ - see \cite{mughal2011phyllotactic, mughal2012dense}. Between these values, the structure is best accommodated by the introduction of a \emph{line slip}, which shears two adjacent spirals with a loss of contacts, as shown in Fig (\ref{bubblespheres}). 
These features were first identified by Pickett \textit{et al.} \cite{Pickett2000}, but not termed ``line slip''.

We have become accustomed to thinking of line slips as being a property of \emph{hard} spheres, and therefore of limited relevance to real physical systems. That point of view is reconsidered here: we experimentally demonstrate the existence of line-slip arrangements in foams with high liquid fraction (see section \ref{sec:experiment}). These observations constitute the first conclusive experimental evidence of such structures (discounting the trivial case of packing ball-bearings in tubes \cite{mughal2012dense}). Furthermore, our experiments with foams demonstrate that line-slip structures can be stable in \emph{soft} systems, a hitherto unexpected outcome. The results presented below improve on our previous attempts with small bubbles in capillaries under gravity, which proved difficult and in which line slips were not clearly seen \cite{meagher2015experimental}.

Our work is stimulated in part by the observation of line slips (albeit rather indistinctly) in some simulations, which use points interacting by (the relatively complex) Lennard-Jones type potentials \cite{wood2013self, douglass2017stabilization}. Here we numerically investigate the stability of structures formed by soft (elastic) repelling spheres in cylindrical channels subject to an applied pressure. It has often been noted that the attractive tail of a pair potential acts rather like an applied pressure, so our simple model should have some generality in qualitative terms and could be compared fruitfully with these earlier simulations.

Our approach enables us to partially map out a rich phase diagram, consisting of a sequence of continuous and discontinuous transitions between uniform structures and line-slip arrangements. We show that in such soft systems line slips gradually vanish with an increasing pressure, while at high pressures only uniform structures remain stable. While it may seem obvious that line slips must disappear at higher pressures, the true scenario is complex, as we shall see.

The paper is organised as follows. In section II we describe our numerical model and present the computed phase diagram. This is compared with our experimental results in section III (for further experimental details see the appendix). The discussion of these results is in section IV, followed by a brief conclusion in section V.

\section{Numerical Model and Results}

\begin{figure}[t]
\begin{center}
\includegraphics[width=\columnwidth ]{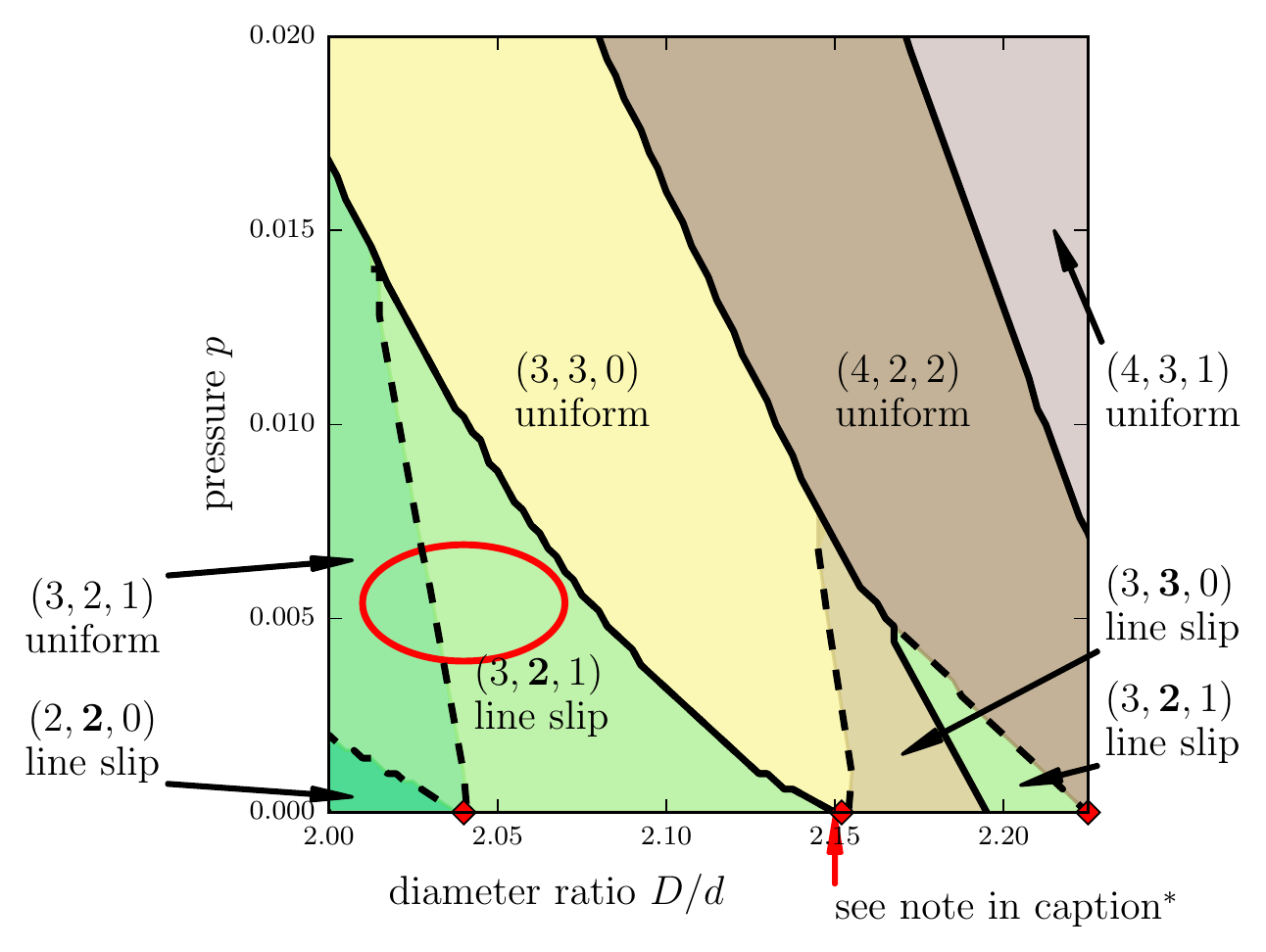}
\caption{
The form of the phase diagram for the range $2.00 \le D / d \le 2.22$ and dimensionless pressures $p \le 0.02$.
Seven different structures are in the plotted range.
With increasing pressure the widths of the line slips decrease and end in triple points.
Discontinuous transitions are displayed as solid black lines and continuous transitions are represented as dashed lines.
The diamond symbols at $p = 0$ correspond to the hard sphere symmetric structures $(3,2,1)$, $(3,3,0)$ and $(4,2,2)$ \cite{mughal2012dense}.
Accounting for the possible metastability (see below) of the observed structure, a rough estimate of the experimentally observed line-slip structure is indicated by the ellipse on this phase diagram. \\
$^*$Note that there is also a very small $(\bm{3}, 2, 1)$ line slip region just below $D / d = 2.15$ \cite{mughal2012dense}, not resolved in the present simulation.
}
\label{phasediagram}
\end{center}
\end{figure}

\begin{center}
\begin{table}[ht]
\captionsetup{justification=centering, font=small }
\caption{Hard Sphere Packings}
 \begin{tabular}{l | p{1.25cm} p{2.5cm}}
 \toprule
 Range & Notation & Description\\ \colrule \colrule
\rowcolor{Gray1}
  $D/d = 2.039$  & (3,2,1)  & uniform \\ \colrule
\rowcolor{Gray2}
  $\;\;\;\;2.039 \leq D/d \leq 2.1413$   & (3,{\bf 2},1)  & {\bf line slip}\\ \toprule
\rowcolor{Gray2}
  $\;\;\;\;2.1413\leq D/d < 2.1545$   & ({\bf 3},2,1)  & {\bf line slip}\\ \colrule
\rowcolor{Gray1}
  $D/d=2.1547$  & (3,3,0)  & uniform \\ \colrule
 \rowcolor{Gray2}
  $\;\;\;\;2.1547< D/d \leq 2.1949$  & (3,{\bf 3},0)  & {\bf line slip} \\ \toprule
 \rowcolor{Gray2}
  $\;\;\;\;2.1949\leq D/d \leq 2.2247$  & (3,{\bf 2},1) & {\bf line slip} \\ \colrule
\rowcolor{Gray1}
  $D/d = 2.2247$  & (4,2,2) & uniform \\ \botrule
\end{tabular}
\captionsetup{justification=centerlast, font=footnotesize}
\bigskip
\caption*{Partial sequence of densest uniform hard sphere packings, together with the line-slip structure, into which they may be transformed (i.e. $p=0$), adapted from Table I of \cite{mughal2012dense}.}
\label{exp:vf_table}
\end{table}
 \end{center}

We begin by describing the model system. We adopt an elementary approach that has proved useful for the description of foams and emulsions \cite{durian1995foam}.
It consists of spheres of diameter $d$, whose overlap $\delta_{ij}$ leads to an increase in energy according to
\begin{equation}
E^S_{ij} = \frac{1}{2} k \delta_{ij}^2\,.
\end{equation}
Here the overlap between two spheres $i$ and $j$ is defined as $\delta_{ij} = \vert \vec{r}_i - \vec{r}_j \vert - d$, where $\vec{r}_i=(r_i, \theta_i, z_i)$ and $\vec{r}_j=(r_j,\theta_j, z_j)$ are the centers of two contacting spheres in cylindrical polar coordinates. A similar harmonic energy term $E^B_i= k / 2 ((D/2 - r_i)-d/2)^2$ accounts for the overlap between the $i$th sphere and the cylindrical boundary.
The spring constant $k$ determines here the softness of the spheres.
We conduct simulations using a simulation cell of length $L$ and volume $V=\pi(D/2)^2L$. On both ends of the simulation cell we impose \emph{twisted} periodic boundary conditions (see \cite{mughal2012dense} for details).

Stable structures are found by minimizing the enthalpy $H = E + PV$ for a system of $N$ soft spheres in the unit cell, where $E = E^S + E^B$ is the internal energy due to overlaps as described before and $P$ the pressure.

The natural units of this simulation are the spring constant $k$ and the sphere diameter $d$.
Therefore, enthalpy and pressure have to be rescaled accordingly to obtain non-dimensional quantities.
In the following we will use the dimensionless enthalpy $h = H / (kd^2)$ and dimensionless pressure $p = P / (k / d)$.

A part of the resulting phase diagram is shown in  Fig \ref{phasediagram}, where for a given value of the diameter ratio $D/d$ and pressure $p$ the enthalpy is minimised by varying the sphere centres, as well as the twist angle and the volume of the simulation cell. This is done for several values of $N$ and the structure with the lowest enthalpy is chosen.
 
For low pressures the minimization was performed with a very general search algorithm (similar to that used previously for the densest packing of hard spheres \cite{mughal2012dense}). We found that the results from these preliminary simulations could be used as initial guesses for a much simpler code (based on conjugate gradient techniques) to map out the higher pressure regions of the phase diagram. Starting with an initial structure (with $N = 3, 4, 5$) we steadily increased the pressure and minimized enthalpy. As a further check we also ran the procedure in the orthogonal direction - i.e. we start with a seed structure with a high value of  $D/d$,  keep the pressure constant while reducing $D/d$ in discrete steps and minimize the enthalpy at each step. In either case the structure with the lowest enthalpy for a given value of  $D / d$ and $p$ is given in the phase diagram.

We plot the phase diagram for the range $2.00 \leq D/d \leq 2.22$ and pressures below $p \leq 0.02$, as shown in Fig \ref{phasediagram}. Along the horizontal axis (i.e. $p=0$) the red diamonds indicate the symmetric hard sphere structures $(3,2,1)$, $(3,3,0)$ and $(4,2,2)$. Intermediate between them are the associated line-slip arrangements (see also  Table \ref{exp:vf_table}). The general trend is that increasing the pressure gradually eliminates the line slips which end up in a triple point and are forced into uniform structures. This transformation equates to a $\SIrange{10}{15}{\percent}$ compression of the hard sphere packings, in the range shown.

We now discuss the features of the phase diagram in detail. Continuous transitions, where the structure changes gradually due to the loss or gain of a contact, are shown in Fig \ref{phasediagram} as dashed lines. Discontinuous transitions, where the structure changes abruptly into another, are indicated with a solid line. Transitions between uniform structures and a corresponding line slip are continuous. This is echoed by the sequence of hard sphere packings (i.e. the structures along the horizontal axis) as listed in Table \ref{exp:vf_table}. An explanation for the connections between a symmetric structures and a line slip for the hard sphere packing can be found in Fig 10 and 12 of \cite{mughal2012dense}.

\begin{figure}[t]
\begin{center}
\includegraphics[width=\columnwidth ]{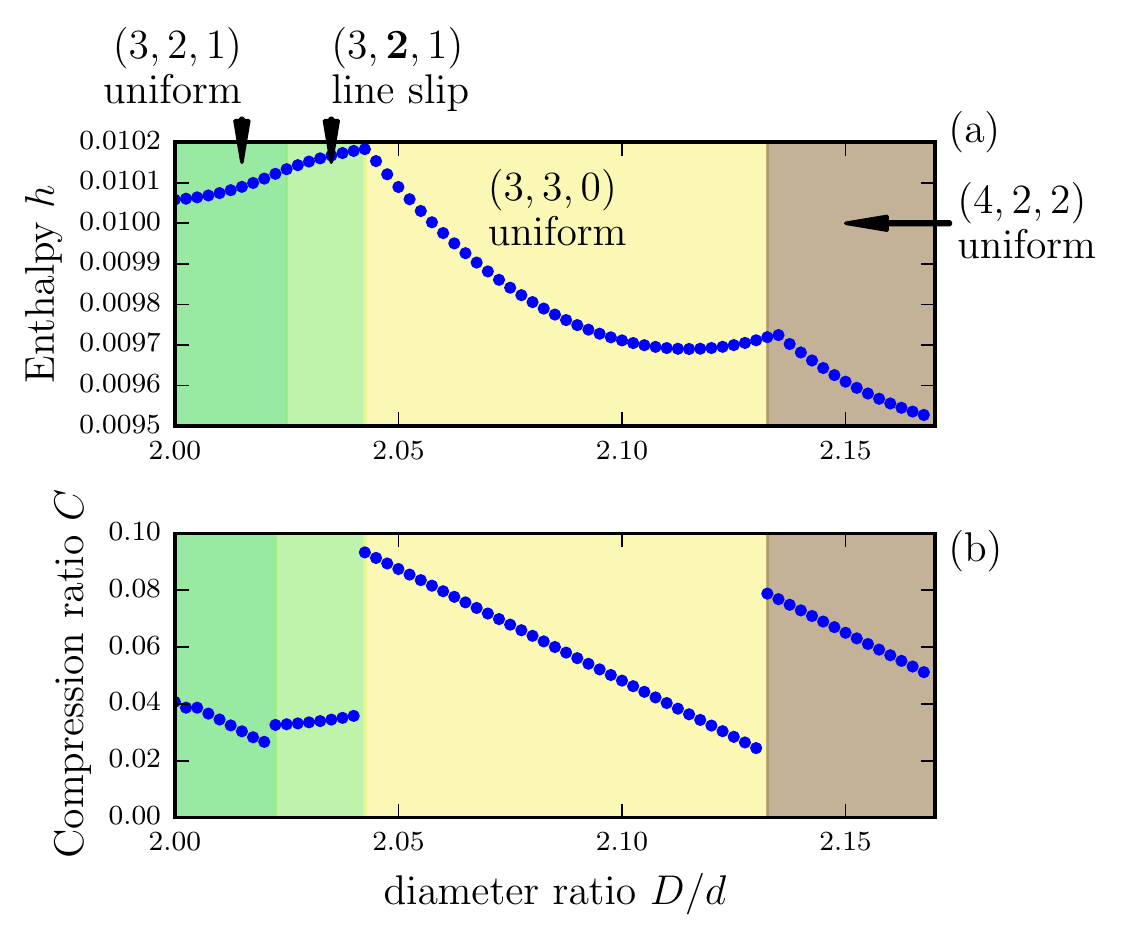}
\caption{
Top: An example of the dimensionless enthalpy $h$ as $D / d$ is varied at a constant pressure $p = 0.01$. The plot shows a horizontal cut through the phase diagram of Fig \ref{phasediagram}. Bottom: Compression ratio $C$ of the soft packings as a function of the diameter ratio $D / d$. Only the transition from $(3,2,1)$ to $(3, {\bf 2}, 1)$ is continuous and does not involve a change in the slope of $H$ at the phase boundary. 
}
\label{enthalpyDd}
\end{center}
\end{figure}

\begin{figure}[t]
\begin{center}
\includegraphics[width=\columnwidth ]{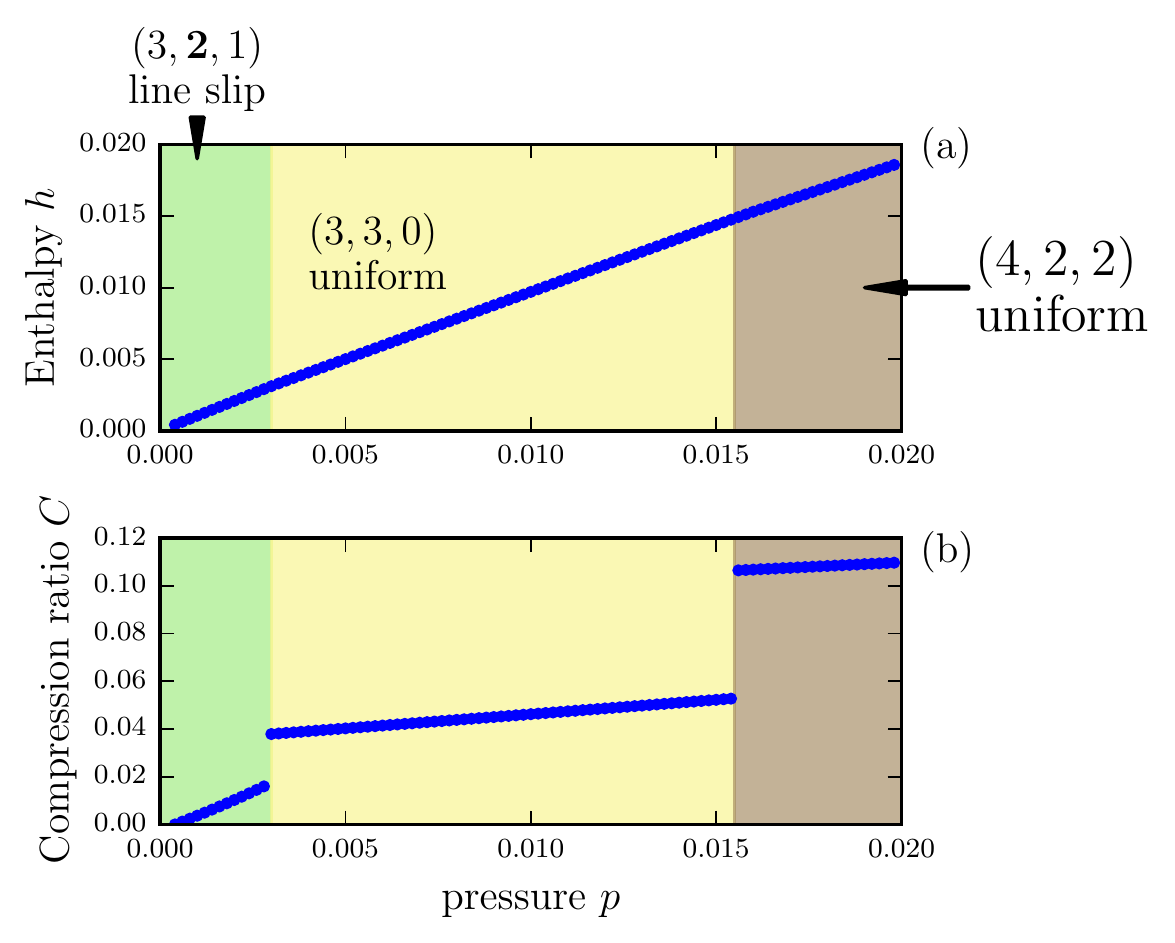}
\caption{
An example of the dimensionless enthalpy $h$ as $p$ is varied at a constant $D/d = 2.1$. The plot shows a vertical cut through the phase diagram of Fig \ref{phasediagram}. Bottom: Compression ratio $C$ of the soft packings as a function of the applied pressure. All the transitions shown are discontinuous, however for the enthalpy the change in the slope at the phase boundaries can only be detected by numerically computing the derivative.
}
\label{enthalpyP}
\end{center}
\end{figure}

A typical example of a discontinuous transition is given by the solid line separating the $(3, \bm{3}, 0)$ and $(3, \bm{2}, 1)$ line-slip regions. We find in the case of soft spheres that with increasing pressure the  $(3, \bm{2}, 1)$ line slip is first to disappear at a triple point followed by the  $(3, \bm{3}, 0)$ line slip at a slightly higher pressure. At still higher pressures only the $(3,3,0)$ and $(4,2,2)$ uniform structures remain stable, separated by a discontinuous phase transition.

On first glance on our phase diagram, the case $(3,2,1) \rightarrow (3,3,0)$  seems to be exceptional, in that only a single line slip, the $(3, \bm{2}, 1)$ line slip, is visible. From Table \ref{exp:vf_table} we would also have expected to see two line slips separating the uniform structures (as in the previous case). However, the  line-slip structure of the $(\bm{3}, 2, 1)$ exists only in the narrow range $2.1413 \leq D/d \leq 2.1545$ for the hard spheres (see  Table \ref{exp:vf_table}) and is not resolved in Fig \ref{phasediagram}.

The distinction between continuous and discontinuous transitions can be illustrated directly via the enthalpy $h$. An example of $h$ in terms of $D / d$ at constant pressure $p = 0.01$ is given in Fig \ref{enthalpyDd}. Continuous transitions, such as $(3, 2, 1)$ uniform (indicated in green) to $(3, \bm{2}, 1)$ line slip (indicated in light green), are not apparent in the variation of $h$. However, discontinuous transitions such as the $(3, \bm{2}, 1)$ line slip to $(3, 3, 0)$ uniform (yellow)  show a change in slope of $h$ at the transition. 

Also shown for comparison in Fig \ref{enthalpyDd} is the compression of the packing subject to the applied pressure. We define the compression as $C=(V_0 - V(p, D / d)) / V_0$, where $V(p, D / d)$ is the volume of the unit cell of the soft sphere packing (for the chosen $p$ and $D / d$) and $V_0$ is the volume of the corresponding hard sphere structure. In the case of uniform structures the volume of the unit cell in the hard sphere case has a unique value \cite{mughal2013screw}. However, for the line-slip structures this is not the case (since the length of the unit cell depends on $D/d$) and instead we compare against the smallest volume of the unit cell for a given hard sphere arrangement of this type \cite{mughal2013screw}.

It is illustrative to also consider the orthogonal trajectory, i.e. variation in the enthalpy $h$ in terms of $p$ at constant diameter ratio. As an example of this we show in Fig \ref{enthalpyP} a vertical cut through Fig \ref{phasediagram}, where we hold the diameter ratio at a constant $D/d = 2.1$ and vary the pressure. The corresponding compression ratio is also shown. The trajectory passes through a transition from $(3, \bm{2}, 1)$ to $(3,3,0)$ and then from $(3,3,0)$ to $(4,2,2)$ - both of these are discontinuous transitions; while the change in the slope cannot be determined by inspection from Fig \ref{enthalpyP} it nevertheless can be clearly observed by taking the derivative of $h$ with respect to $p$ along the trajectory.

In due course we shall use the methods developed here to investigate the phase diagram for higher $D/d$ up to $D / d \leq 2.7379$, beyond which the nature of the hard sphere packings changes \cite{mughal2012dense, fu2016hard}, and to higher pressure.

\section{Experiments}
\label{sec:experiment}

We now describe our experimental procedure and results (for further details of the set-up see the appendix). For a convenient experimental counterpart, we choose to observe structures of columns of bubbles under \emph{forced drainage} that is, a steady input of liquid from above \cite{weaire1993steady, weaire1997review, koehler2000generalized}. In the past this has been extensively studied for columns of bulk foam where $D / d$ was very large. It was found that this results in convective instabilities for flow rates that gave rise to higher liquid fractions, confining experiments on static foam structures to relatively dry foam \cite{Crawford1998}. The corresponding theory was therefore developed using approximations appropriate in the dry limit, that is, in terms of liquid flow in a network of narrow Plateau borders.

The present work leads us to consider columns of wet foams, that is, with the drainage rate high enough to produce near-spherical bubbles.
We find no such instability in the case of the confined columnar structures of large bubbles considered here; 
hence the wet limit can be reached at a certain flow rate.
Below that point, the liquid fraction can be varied, producing ordered columnar structures closely analogous to those described above.

For the experimental set-up \cite{weaire1993steady, hutzler1997moving}, monodisperse bubbles are produced by blowing air through a needle into a commercial surfactant Fairy Liquid. The surfactant solution contains 50\% of glycerol in mass to increase the viscosity, smooth the transition between structures and to allow us observe more easily unstable bubble arrangements. Tuning the gas flow rate ($q_{0}\sim{}1ml/min$) allows us to produce monodisperse bubbles of controlled size ($d\sim{}2.5mm$).

When released into a vertical cylinder, the bubbles are observed to self-organise into ordered structures (we used a glass tube for which the diameter is $5mm$ and the length is $1.5m$). The resulting foam column is put under forced drainage by feeding it with surfactant solution from the top (liquid flow rate $Q$ up to $\sim10ml/min$). We estimate the capillary pressure for the bubbles to be of the order of approximately $10 Pa$, which is an order of magnitude smaller than the hydrostatic pressure - thus the bubbles are in the regime whereby they can easily be deformed and the foam can be treated as a packing of relatively soft objects - see appendix.

\begin{figure}
\begin{center}
\includegraphics[width=0.7\columnwidth, clip=true, trim=0 0 15 0]{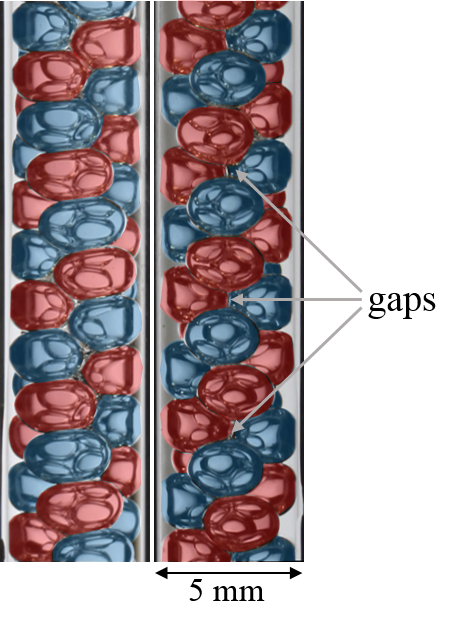}
\caption{Photograph of columnar structures of bubbles under forced drainage, showing about forty bubbles.
Left: $(3,2,1)$ uniform structure occurring at a relatively low flow rate.
Right:  Bubbles adopt the $(3,\bm{2},1)$ line-slip structure at a relatively high flow rate.
It is similar to the line-slip structure of Fig (\ref{bubblespheres}b). The  coloring highlights the two interlocked spirals of the (3,2,1) and $(3,\bm{2},1)$ structures. The ellipse marked on Fig \ref{phasediagram} corresponds roughly to its estimated position on the phase diagram. 
}
\label{bubblespheresx}
\end{center}
\end{figure}

We show in Fig \ref{bubblespheresx} an example of the observation of the $(3, \bm{2}, 1)$ line-slip structure in accord with expectations based on the soft sphere model, described above. Note that the bubbles are distorted into smooth ellipsoids by the flow.
The extent of the loss of contact in the observed line-slip structure is roughly equal to that of the simulated structure at the position marked with an ellipse in the phase diagram. This is also consistent with our estimates of equivalent pressure $p$ and equivalent $D / d$, where the equivalent pressure is determined from the local packing fraction.

While the boundaries plotted in Fig 2 demarcate the borders between the
lowest enthalpy structures, preliminary simulations indicate that
such structures can be metastable. Hence we indicate by an ellipse on
Fig \ref{phasediagram} the possible range over which the experimentally observed
structure can exist (where we also have accounted for possible sources
of experimental error). We are currently in the process of examining
the metastability of soft sphere columnar structures by performing
further experiments and simulations

The shape distortion due to liquid flow, together with that due to the forces between bubbles, and hence the equilibrium structure itself, will be analysed in a subsequent paper, together with a wide range of observations. Those forces are only very crudely modelled in the present paper. The necessary theory of  drainage of stable structures of  wet foam has not previously been formulated.

\section{Discussion}

Considering that our original study of hard-sphere arrangements in a cylinder was motivated as a packing problem \cite{mughal2011phyllotactic}, it may have seemed plausible that the structures described above would only be of importance close to the incompressible athermal limit. Recently, however, these original findings have been found to be relevant in the \emph{self-assembly} of hard spheres in cylindrical channels \cite{fu2016assembly}. Remarkably, it is found that at finite temperature and pressure both uniform and line-slip arrangements can be observed: a steadily increasing pressure can induce transitions between structures, while at a high compression rate some arrangements are found to be dynamically inaccessible  \cite{fu2016assembly}, these results are illustrated in terms of a phase diagram that bears a resemblance to the results presented here. This is due to the fact that the thermally agitated hard spheres have an effective radius that scales with the mean free path, this leads to an effective interaction reminiscent of a soft potential. These results only indirectly hint at the manner in which the hard sphere results are modified in the presence of soft interactions. The present results have the advantage of directly addressing the stability of line-slip arrangements in softly interacting systems.

These results may also be fruitfully compared with previous studies employing the Lennard-Jones potential, where both uniform structures and line-slip arrangements are observed \cite{wood2013self}. On the other hand, in the case of very soft potentials, such as the Yukawa interaction, line-slips have not been either predicted to exist \cite{oguz} or indeed observed in experiments \cite{wu}. Determining the robustness of the phase diagram presented in Fig \ref{phasediagram} as a function of the hardness/softness of the interaction potential remains an open question

Another related study is that of Rivier and Boltenhagen \cite{Boltenhagen1996} who observed structural transitions in cylindrical \emph{dry} foam under pressure, well outside of the range of our investigation.

\section{Conclusions}

In conclusion, questioning the relevance of hard-sphere properties to physical systems, consisting of \emph{soft} spheres has led us into a new territory, not previously explored.
We have shown that line-slip arrangements are a feature of soft systems (even in the absence of thermal agitation), thus extending their usefulness to encompass a range of commonly encountered substances including dusty plasmas, foams, emulsions and colloids. It is of direct relevance to some physical systems (foams, emulsions) and offers qualitative interpretation in others.
It extends into another dimension (that of pressure) the elaborate table of hard-sphere structures previously found.

These results are strictly only relevant to a microscopic system which is in thermal equilibrium. For some systems, such as those currently under investigation \cite{tsytovich2007plasma, zerrouki2008chiral, yin2003self, oguz, fu2016assembly, wu, Mickelson2003, Li2005, Lohr2010}, many observed structures will be metastable (i.e. not structures of lowest enthalphy). We therefore are engaging in complementing the work of this paper by exploring the wider context of metastable structures and transitions between them. This will establish a full comparison with various experiments (also on the way). We will furthermore in due course also use the methods developed here to investigate the phase diagram for higher $D/d$ up to $D/d \leq 2.7379$, beyond which the nature of the hard sphere packings changes \cite{mughal2012dense, fu2016hard}, and to higher pressure.

Finally, it is worth noting that there is considerable current interest in columnar crystals, of the type described here. Applications include various helical biological microstructures, such as viruses \cite{erickson1973tubular} and bacteria \cite{amir2012dislocation}. In addition, there are a range  of problems involving the self assembly (or packing) of spheres and particles in tubes \cite{oguz, fu2016assembly, wu, Mickelson2003, Li2005, Lohr2010}. The context of this work is further expanded by the recently demonstrated analogy between line-slip structures and dislocations in the crystalline phase on a cylinder \cite{beller2016}.
 
\acknowledgments
A. M. acknowledges support from the Aberystwyth University Research Fund.
This research was supported in part by a research grant from Science Foundation Ireland (SFI) under grant number 13/IA/1926 and from an Irish Research Council Postgraduate Scholarship (project ID GOIPG/2015/1998). We also acknowledge the support of the MPNS COST Actions MP1106 `Smart and green interfaces' and MP1305 `Flowing matter' and the European Space Agency ESA MAP Metalfoam (AO-99-075) and Soft Matter Dynamics (contract: 4000115113).

\section*{Appendix} 
Here we briefly discuss the experimental set-up used to produce the columnar arrangement of bubbles. We begin by describing a method for generating monodisperse foams in a column. We then discuss how the liquid (or packing) fraction and its effect on the morphology of the foam can be controlled.

\begin{figure*}[h]
a) \includegraphics[scale=0.35]{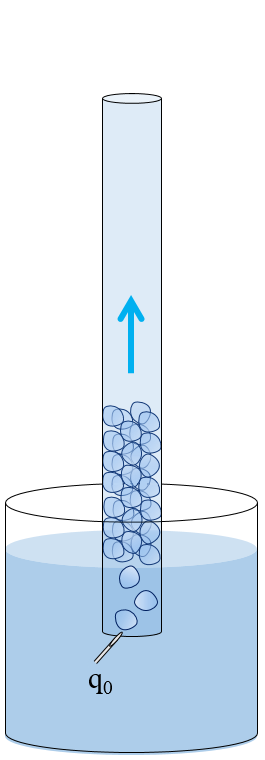}
\hspace{4.4cm}
b) \includegraphics[scale=0.35]{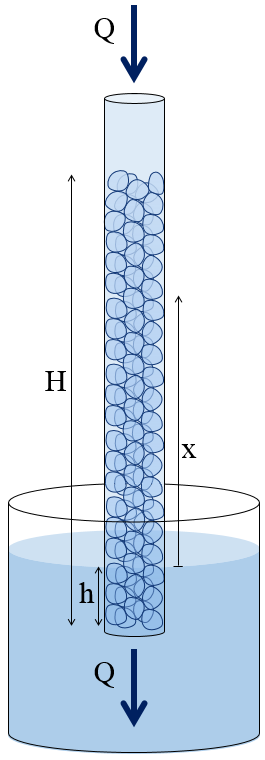}
\caption{Set-up for producing columnar foam structures. a) Gas is introduced at a constant flow rate $q_0$ into a cylindrical column b) Liquid is added at a constant flow rate $Q$ at the top of the column (force drainage).
$H$ is the height of the entire foam column, $h$ is the height of the submerged foam and $x$ measures the distance to the liquid pool.}
\label{setup}
\end{figure*}

As shown in Fig (\ref{setup}a), a steady stream of monodisperse bubbles is produced by blowing air through a needle dipped in a surfactant solution. Tuning the gas flow rate ($q_{0}\sim{}1ml/min$) allows us to produce monodisperse bubbles for a range of diameters.
The bubbles are collected in a cylindrical tube, where they form a columnar foam.

The bubbles are approximately of diameter $d\sim{}2.5\ mm$ with a standard deviation of $0.04\ mm$. The bubble diameter is determined by squeezing a small amount of foam between two horizontal, parallel plates, separated by a controlled gap to obtain a monolayer of bubbles and image analysis of the top view.

For a bubble in our experiments the capillary pressure is 
\[
p_{\mathrm{c}} = \gamma / r,
\]
where $\gamma=0.03 Nm^{-1}$ is the surface tension and $r=d/2$ is the radius of the bubble. The hydrostatic pressure, 
\[
p_{\mathrm{h}}=\rho g x,
\]
for a given bubble depends on the height $x$ of the bubble in the column, as measured from the liquid/air interface (see Fig (\ref{setup}b)), as well as the density of the liquid $\rho$ and gravity $g$. By balancing $p_{\mathrm{c}}$ with $p_{\mathrm{h}}$, $x$ can be determined to $x = \gamma / (p \rho r) \approx 1mm$.
The column used for experiments is approximately $1.5m$. Thus, we can safely assume that the hydrostatic pressure dominates in the experiments and the foam can be regarded as \emph{soft} packing of bubbles.

The resulting foam column is put under forced drainage by feeding it with surfactant solution from the top with a liquid flow rate $Q$ up to $Q\sim10ml/min$.  As shown in Fig (\ref{setup}b), the relevant experimental parameters are the height of the foam $H$, the total depth $h$ at which it is immersed in the surfactant solution and $x$ the position in the foam (which is taken from the liquid/air interface). By Archimedes principle the average packing fraction of the foam is given by $\bar{\phi}=1 - h/H$.  

\begin{figure*}[h]
a) \includegraphics[scale=0.31]{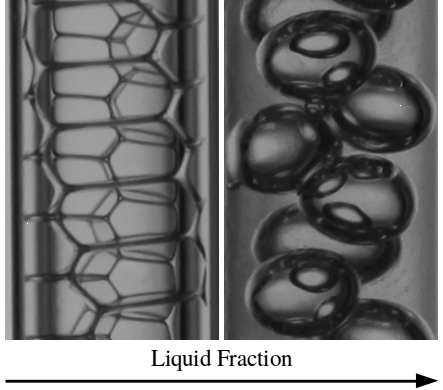}
\hspace{1cm}
b) \includegraphics[scale=0.31]{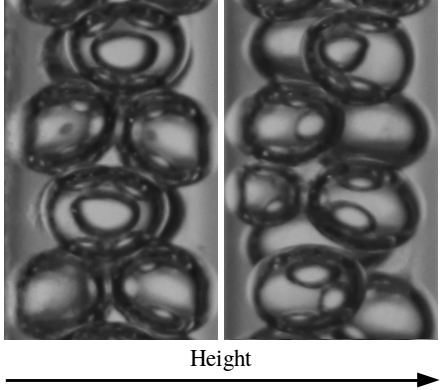}
\caption{
Examples of columnar structures in foams.
(a) shows the same section of the tube for two different flow rates, corresponding to $\bar{\phi}=0.03$, $(3, 2, 1)$ structure on the right, and $\bar{\phi} = 0.5$, $(2, \bm{2}, 0)$ structure on the left.
(b) shows a column at two different vertical positions, $x = 15\ cm$ in the $(2, 2, 0)$ structure on the left and $x = 30\ cm$ in the $(3, \bf{2}, 1)$ structure on the right.
The bubble size is $d \sim{} 2.5\ mm$.
For more details see also table \ref{table:appendix}.
}

\label{photos}				
\end{figure*}

In addition to $\bar{\phi}$, the local packing fractions $\phi(x)$ in a section $\Delta x$ is determined from the photographs of the structure (see Figure \ref{photos}). The diameter of the tube $D$, the diameter of the bubbles $d$. Hence, we have
\begin{equation}
\phi(x)=\left[\frac{N(x)\pi{}d^3}{6}\right]\times\left[\frac{4}{\pi{}D^2 \Delta x}\right],
\end{equation}  
where $N$ is the number of bubbles observed in considered section of the tube. 

We find that the type of structure observed in the column depends strongly on $\phi(x)$.
An incomplete list of observed structures as a function of packing fraction are shown in table \ref{table:appendix};
examples of foam structures are illustrated in Fig. \ref{photos}.

In order to roughly indicate the location of a structure in the $p-D/d$ phase diagram of Fig \ref{phasediagram} we proceed as follows. The ratio $D/d$ is computed using the tube and bubble diameters respectively. The local packing fraction (as defined in the appendix) from the experiment is directly comparable to the packing fraction $\phi$ in the simulation, which depends on $p$. Thus by comparing experimental results with our simulations we are able to compute a value of $p$ for a given value of experimentally determined $\phi(x)$.

\begin{center}
\begin{table}[ht]
\captionsetup{justification=centering, font=small}
\caption{Partial list of experimentally observed structures}
\begin{tabular}{c c r r c} 
\toprule
 Structure & $x$ ($cm$) & $\bar{\phi}$ & $\phi(x)$ & $Q\ (ml/min)$ \\ [0.5ex] \colrule\colrule
 \rowcolor{Gray1}
 (3,2,1) & 15 & $\sim{}0.97$ & $\sim{}0.97$ & $\sim{}0$ \\ \colrule
  \rowcolor{Gray2}
 (2,\textbf{2},0) & 15 & $\sim{}0.50$ & $\sim{}0.38$ & $\sim{}10$ \\ \colrule
  \rowcolor{Gray1}
 (2,2,0) & 15 & $\sim{}0.48$ & $\sim{}0.46$ & $\sim{}2$ \\ \colrule
  \rowcolor{Gray2}
 (3,\textbf{2},1) & 30 & $\sim{}0.48$ & $\sim{}0.48$ & $\sim{}2$ \\ \colrule
\end{tabular}
\captionsetup{justification=centering, font=footnotesize}
\bigskip
\caption*{Partial list of experimentally observed structures as a function of local and averaged packing fraction as well as height $x$.}
\label{table:appendix}
\end{table}
\end{center}

\newpage

\bibliographystyle{nonspacebib}


\end{document}